\documentclass[11pt]{article}
\usepackage{epsfig}
\newcommand{\sss}{\vspace{.2in}}

\newcommand{\be}{\begin{equation}}
\newcommand{\ee}{\end{equation}}
\newcommand{\bea}{\begin{eqnarray}}
\newcommand{\eea}{\end{eqnarray}}
\newcommand{\sn}{{\rm sn}}
\newcommand{\cn}{{\rm cn}}
\newcommand{\dn}{{\rm dn}}

\begin{document}
~\hfill{\footnotesize UICHEP-TH/01-2, IP/BBSR/01-5,~~\today}
\sss
\sss
\begin{center}
{\Large {\Large \bf Some Exact Results for Mid-Band and Zero Band-Gap}}
{\Large {\Large \bf States of Associated Lam\'{e} Potentials}}
\end{center}
\vspace{.5in}
\begin{center}
{\large{\bf
   \mbox{Avinash Khare}$^{a,}$\footnote{khare@iopb.res.in} and
   \mbox{Uday Sukhatme}$^{b,}$\footnote{sukhatme@uic.edu}
 }}
\end{center}
\vspace{.6in}
\noindent
a) \hspace*{.2in}
Institute of Physics, Sachivalaya Marg, Bhubaneswar 751005, Orissa, India\\
b) \hspace*{.2in}
Department of Physics, University of Illinois at Chicago, Chicago, IL 60607-7059, U.S.A. \\
\sss
\sss
\begin{abstract}
Applying certain known theorems about one-dimensional periodic 
potentials, we show that the energy spectrum of   
the associated Lam\'{e} potentials 
$$a(a+1)m~{\rm sn}^2(x,m)+b(b+1)m~{\rm cn}^2(x,m)/{\rm dn}^2(x,m)$$ 
consists of a finite number of bound bands followed by a continuum band when
both $a$ and $b$ take integer values. 
Further, if $a$ and $b$ are unequal integers, we show that there must exist 
some 
zero band-gap states, i.e. doubly degenerate states with the same number of
nodes. More generally, in
case $a$ and $b$ are not
integers, but either $a + b$ or $a - b$ 
is an integer ($a \ne b$), we again show that several of the 
band-gaps vanish due to degeneracy of states with the same number
of nodes.
Finally, when either $a$ or $b$ is an integer and the other 
takes a half-integral
value, we obtain exact analytic solutions for  
several mid-band states.  
\end{abstract}
\newpage

\sss
{\noindent\bf I. Introduction} 

The energy spectrum of electrons on a lattice is of central importance in 
condensed matter physics. In particular, the knowledge of the existence and 
locations of band edges and band gaps determines many physical properties.  
Unfortunately,
even in one dimension, there are very few analytically solvable periodic 
potential problems in quantum mechanics. The Lam\'{e} potential
\be\label{1}
V(x) = a(a+1)m~{\rm sn}^2(x,m) ~,~ a =1,2,3,... ~~,
\ee
is well-known to be exactly solvable. Here ${\rm sn}(x,m)$ is a Jacobi elliptic function of real elliptic modulus
parameter $m$ $( 0\leq m\leq 1)$ with period $4K(m)$. 
For simplicity, from 
now onward, we will not explicitly display the modulus parameter $m$ as an 
argument of Jacobi elliptic functions \cite{gr}. 

Recently, we have vastly extended \cite{ks} the list of known solvable potentials by exploiting  two different directions.
Firstly, we have shown that the supersymmetric partners of the Lam\'{e} potential
constitute a
wide class of new exactly solvable periodic potentials, which are distinctly different from the Lam\'{e} potential of eq. (1) (for $a>1$), even
though they have the same energy band structure. 
Secondly, we have shown \cite{ks} that the associated Lam\'{e} potentials \cite{mw}
\be\label{2}
V(x) = pm~{\rm sn}^2 \, x + qm~{{\rm cn}^2 \,x\over {\rm dn}^2 \,x}~,~ 
~~p \equiv a(a+1)~,~~q \equiv b(b+1)~~,
\ee
(which constitute a much richer class of periodic potentials) 
and their supersymmetric partners yield many additional solvable and
quasi-exactly solvable (QES) periodic problems provided $a + b$ and/or 
$a - b$ is an integer.
Here, like ${\rm sn}\,x$, the Jacobi elliptic functions ${\rm cn}\,x$ and 
${\rm dn}\,x$ also have the same modulus parameter $m$ which, for 
notational convenience, is not explicitly displayed. Without any loss of
generality, we shall always consider associated Lam\'{e} potentials with $p \ge q $ \cite{ks}.
All associated Lam\'{e} potentials have a period $2K(m)$, but the special 
case $p=q$ has a period $K(m)$. 

There are several issues which were not addressed in our previous paper \cite{ks}, hereafter simply referred to as I. 
For example, in the case of the Lam\'{e} potential (1), it is well known that 
there are $a$ ($\mid a+1 \mid $) bound bands followed by a continuum band when 
$a$ is a 
positive (negative) integer \cite{ar}. 
What about the associated Lam\'{e} case, 
especially when both $a$ and $b$ are integers? In particular, for these 
cases, are there
only a finite number of bound bands followed by a continuum band? What 
happens if $a$ and $b$ are not integers but either $a+b$ or $a - b$ are integers?
Furthermore, for the Lam\'{e} potential, when $a$ takes half integral
values (say $a= n+1/2,~~n=0,1,2,...$), then $(n+1)$ doubly-degenerate 
solutions are known 
\cite {mw,ar} for 
the mid-band states (i.e. states with period $8K$). Can one  
also analytically obtain the mid-band 
states for the associated Lam\'{e} potentials? 

The purpose of this paper is to address the issues raised above. 
In particular,
using certain known theorems about periodic problems in one dimension,
we show that even in the associated Lam\'{e} case, there are only a finite
number of bound bands followed by a continuum band in case both $a$ and $b$
are integers. Further, so long as $a$ and $b$ (and hence $p,q$) are unequal 
integers, 
then we also show that some of the band gaps (either of period $2K$ or $4K$)
disappear.
On the other hand, if $a,b$ are not integers but either $a+b$ 
or $a - b$ is an 
integer ($a \ne b$), then in general there are an infinite number of bands 
out of which, but for the lowest few, all other band gaps of either period 
$2K$ or $4K$ vanish.
Finally, when $a$ is a half integer and $b$ is an integer, we obtain
several exact mid-band states (i.e. states of period $8K$).

The plan of this paper is as follows. In Sec. II we describe some known
theorems about the number of band gaps in a periodic potential in one 
dimension.
Using these theorems, we show that if $a$ and $b$ are integers 
such that $p > q >0$, 
then the associated Lam\'{e} potential has also only $a$ bound bands followed by
a continuum band. Further, 
if $a - b$ is an even (odd) integer, 
then there are $b$ doubly degenerate band
edges of period $4K$ ($2K$) (i.e. in these cases the corresponding band gaps
are zero). Unfortunately, we are unable to obtain energy eigenstates for 
any of these $2b$ states. However, we do obtain exact expressions for the 
remaining $2a+1$ band edges. In particular, if $a - b$ is an even (odd) 
integer, then
one can obtain the energy eigenvalues and eigenfunctions for the $a+b+1$ 
($a - b$) states of period $2K$ and $a - b$ ($a+b+1$) states of period $4K$.
In Sec. III we discuss the case when both $a,b$ are half integral  
and such that $p>q>0$. In this case we show that if   
$a - b$ is an even (odd) integer, 
one can obtain exact eigenvalues and eigenfunctions 
for the $b+1/2$ doubly degenerate states of period $4K$ ($2K$) (i.e. in 
these cases, the corresponding band gaps are zero) as well as
$a - b$ nondegenerate states of the same period. Unfortunately, in this case
one is not able to obtain any eigenstates of period $2K$ ($4K$).
In Sec. IV we consider the case when $a$ is half-integral ($a=k+1/2$) and
$b$ is an integer ($b=s,~~s=0,1,2,...,N,~~k=N-s$) and show that for every 
possible value of $s$, one can 
obtain exact, doubly-degenerate,   
$k+1$ mid-band states of period $8K$.
Finally, in the last section we summarize the results obtained in this paper
and point out some open problems.

\sss
{\noindent\bf II. $a,b$ Integral and Finite Number of Bound Bands} 

Consider the Schr\"odinger equation for the associated
Lam\'{e} potential specified in eq. (2) for the usual case of a particle of mass 1/2 using units with $\hbar=1$:
\be\label{2.1}
- {d^2\psi (n)\over dx^2} + \left[ a(a+1)m \sn^2 (a) +b(b+1) m
{\cn^2\,x\over \dn^2\,x}\right] \psi (x) = E\psi (x)~.
\ee
Following the procedure described in I, we substitute 
\be\label{2.2}
\psi (x) = [ \dn \,x]^{-b} y(x)~,
\ee
yielding
\be\label{2.1a}
y''(x) +2bm \frac{\sn \,x~\cn \,x}{\dn \,x} y'(x)
+[\lambda -(a+b)(a+1-b)m\sn^2 \,x]y(x) = 0~.
\ee
On further substituting $\sn \,x \equiv \sin t, y(x)\equiv z(t)$,
it is easily shown that $z(t)$ satisfies Ince's equation
\be\label{2.3}
(1+A \cos 2t) z"(t) + B \sin 2t z'(t)+(C+D \cos 2t) z(t) =0~,
\ee
where
\bea\label{2.4}
A & = & {m\over 2-m}~, \ ~~B = {(2b-1)m\over 2-m}~ , \ ~~C =
{\lambda -(a+b)(a+1-b)m\over 2-m}~, \nonumber \\ 
D & = & {(a+1-b)(a+b)m\over
2-m}~, ~~~~~\lambda = E - mb^2~.
\eea
Now several exact results for Ince's equation are known in the literature. In particular, it is 
known that the system satisfying Ince's eq.
(\ref{2.3}) has {\it at most} $j+1$ band gaps of period $\pi
[2\pi]$ in case the polynomial $Q(\mu) [Q^*(\mu)]$ has 
nonnegative integral roots, the highest of which is $j$
\cite{mw1}. Here the quadratic polynomials $Q(\mu)$ and
$Q^*(\mu)$ are given by 
\be\label{2.5}
Q(\mu) = 2A\mu^2 - B\mu - {D\over 2}~,~~~
Q^* (\mu) = 2A(\mu-{1\over 2})^2 -B(\mu-{1\over 2}) -
{D\over 2}~.
\ee
On the other hand, if $Q(\mu) [ Q^*(\mu)]$ has
negative integral roots, the smallest of which is 
$-j_0-1$, then the system satisfying Ince's eq.
(\ref{2.3}) has {\it at most} $j_0+1$ band gaps of period
$\pi [2\pi]$. It must be noted here that in the notation
of \cite{mw1}, the lowest band
gap of period $\pi$ {\it always} exists and is from $E =
- \infty$ to $E=E_0$, when $E_0$ denotes the energy of the
lower edge of the lowest energy band.

Using these theorems it is easily shown \cite{mw} that the
Lam\'{e} potential (1) has $a \,(\mid a+1 \mid)$ 
bound bands followed by a continuum band in case $a$ is a positive (negative) integer. 

Let us now apply these results to the associated Lam\'{e}
potential (2). We might add here that the period $\pi [2\pi]$
for Ince's equation corresponds to period $2K(m)[4K(m)]$ in
the associated Lam\'{e} case (note $\sn \,x \equiv \sin t)$. On
using the expressions for A,B,C,D as given in
eq. (\ref{2.4}) it is easily shown that the roots of
$Q(\mu)$ are at
\be\label{2.6a}
\mu_1 = {a+b\over 2}~ ,~~ \mu_2 = {b-a-1\over 2}~,
\ee
while the roots of $Q^*(\mu)$ are at
\be\label{2.7}
\mu^*_1 = {a+b+1\over 2}~, \ \ \mu^*_2 = {b-a\over 2}~.
\ee
These roots are integral if and only if a+b and/or a - b take
integer values. In this section we consider the case when
both a and b are integral while in the next section we
consider the other possibilities. 

When both a and b take integer values then it follows from
eqs. (\ref{2.6a}) and (\ref{2.7}) that either the roots
$\mu_1$ and $\mu^*_2$ or the roots $\mu_2$ and $\mu^*_1$ are
integral. In particular, it both a and b or odd or even
then the roots $\mu_1$ and $\mu^*_2$ are integral while if
one of them is odd and the other even then $\mu_2$ and
$\mu^*_1$ are integral. Thus in both cases, there are only
finite number of band gaps of period $\pi$ and $2\pi$ ( and
hence of periods $2K$ and $4K$ for the associated Lam\'{e}
potential (2)). Hence it follows that when both $a$ and $b$
take integer values, then there are only a finite number of
bound bands followed by a continuous band.

In fact, as we show now, these theorems when supplemented
with the exact results obtained in I (see Table III of I)
tell us quite a lot about the nature of band structure in
these cases. We consider the two cases of $a - b$ being odd or
even integer separately. 

(i) $a - b$ = odd integer:

{}From eq. (\ref{2.6a}) it then follows that there are at most
${a - b+1\over 2}$ number of band gaps of period $2K$.
An examination of the few explicit
cases confirms the fact that there are indeed so many band
gaps (and not less) of period $2K$. This implies that in this
case there are only $a - b$ number of nondegenerate states of
period $2K$. Quite remarkably, all these $a - b$ states are QES
states. In particular, the solution for these states can be
obtained from Table III of I in case p = a(a+1) while $q =
[a-(a - b-1) ] [ a-(a - b)]$. 

On the other hand, from eq. ({\ref{2.7}), it follows that in
case $a - b$ is an odd integer, then there are at most ${a+b+3\over
2}$ number of band gaps of period $4K$. 
However, specific examples
show that there are in fact only ${a+b+1\over 2}$ number of band gaps.
This also follows from eq. (\ref{2.7}) in case we take $a>0$ but instead
of $b$ take $-b-1$ (note that $q$ is invariant under $b \rightarrow -b-1$).
We see that $\mu_2^{*}$ has an integral root at $-{{a+b+1}\over 2}$
and hence there are at most ${a+b+1\over 2}$ number of band gaps 
i.e. $a+b+1$ number of nondegenerate states of period $4K$. Again
all these are QES states, the solution for which is obtained
from Table III of I in case $p = a(a+1), q = [a-(a+b)][ a-(a+b+1)]$. 

Thus we see that when $a - b$ is an odd integer then there are
${a+b+1\over 2}$ band gaps of period $4K$ but only ${a - b+1\over 2}$ 
band gaps of period $2K$ and the corresponding band edges are known
in principle from Table III of I. However, since the band edge
wavefunctions arranged in order of increasing energy are of
period $2K, 4K, 4K, 2K, 2K,...$, hence it follows that in
this case there must also be $b$ band gaps of period $2K$ which
must be of zero width i.e., there must be $b$ doubly
degenerate states of period $2K$. Unfortunately, so far, we
have not been able to obtain either the eigenvalues or
the eigenfunctions of even one of these states. Thus in this
case there are in all $a$ bound bands followed by a continuum band 
out of which the top $b$
bound bands are a bit unusual in that both of their band
edges have period $4K$ and two degenerate states of period $2K$
lie inside each of these bound bands.

As an illustration, consider the case of $p = 12, q = 6$, i.e.
$a = 3, b = 2$. From the above discussion it follows that there must be 1 QES band edge of period $2K$  which 
is the ground state. Using
Table III  of I, it is easily seen that the eigenvalue and the
eigenfunction of this state is
\be\label{2.20}
\psi_0 = \dn^3 \, x~,~~E_0 = 9m~. 
\ee 
In addition, there must be 6 nondegenerate QES band edges 
of period $4K$ and the
eigenvalues and eigenfunctions for these six states are
easily obtained. In particular, it is easily shown that three
of the eigenstates have the form
\be\label{2.21}
\psi_{1,6,9} = \frac{\cn \, x}{\dn^2 \, x} [A +B\sn^2 \, x +D\sn^4 \, x]~,
\ee
and the corresponding three eigenvalues $E_{1,6,9}$ satisfy  
the cubic equation
\be\label{2.22}
\lambda^3 -4(8-m)\lambda^2 +48(4+m)\lambda -576m =0~,~~E=\lambda +1+4m~.
\ee
The other three eigenstates have the form
\be\label{2.23}
\psi_{2,5,10} = \frac{\sn \, x}{\dn^2 \, x} [A +B\sn^2 \, x +D\sn^4 \, x]~,
\ee
and the corresponding three eigenvalues $E_{2,5,10}$ satisfy  
the cubic equation
\be\label{2.24}
\lambda^3 -8(4+m)\lambda^2 +48(4+7m)\lambda -576m(3+m) =0~,~~E=\lambda +1+m~.
\ee
In view of the oscillation theorem, it is then clear that 
there must be a pair of doubly degenerate states ($\psi_{3,4}$)
and ($\psi_{7,8}$) of period $2K$ whose energy must go to
4(14) and 16(17) respectively as $m \rightarrow 0(1)$. Thus
whereas four states $\psi_{2,3,4,5}$ must merge at $E=14$
as $m \rightarrow 1$ the other four states $\psi_{6,7,8,9}$ 
must merge at $E=17$ as $m \rightarrow 1$. Thus, as shown in
Fig. 1, in this case there are three bound bands followed by
a continuum band. The two upper-most bound bands have both
of their band edges of period $4K$ and in between are the 
pair of doubly degenerate states $\psi_{3,4}$ and $\psi_{7,8}$
of period $2K$ whose energy eigenvalues and eigenfunctions are
not known analytically. We have therefore computed these 
energy eigenvalues numerically and these are shown by the dotted lines
in the figure.                                                                                        
\begin{figure}[ht] \label{fig1}
    \centering
    \epsfig{file=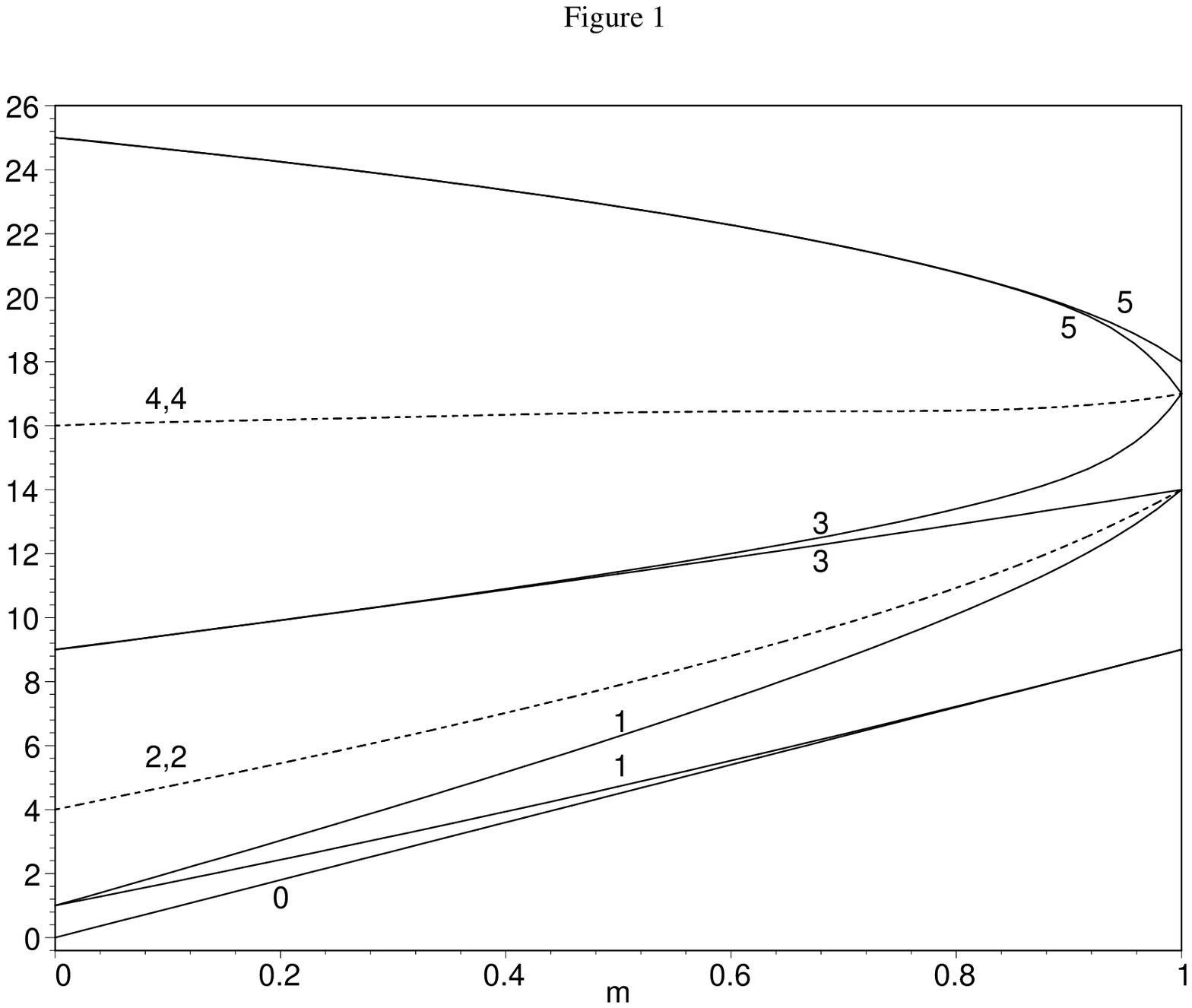,clip=,width=5.0in}
\caption{Band edge energies for the associated Lam\'e potential (12,6)
as a function of the elliptic modulus parameter $m$. The band edges
are labelled by the number of wave function nodes in the interval
$2K(m)$. Note that the band gap between the two states with 2 nodes
as well as with four nodes is zero, that is $E_3=E_4$ and $E_7=E_8$. 
The energy eigenvalues of these degenerate
states have been calculated numerically and is shown by dotted lines. }
\end{figure}

As another illustration, consider the case of $p = 6, q = 2$, i.e.
$a = 2, b = 1$. From the above discussion it follows that in
this case there must be 1 QES band edge of period $2K$ and 4
QES band edges of period $4K$ and interestingly enough the
eigenvalues and eigenfunctions for these five states were
already given in the Table IV of I. Further, as stated
in I, two states of period $2K$ must also be present but so
far we are unable to obtain these states analytically. 
However, what was
not clear at that time was that the two 
states  of period $2K$ must be degenerate and whose energy goes to 4~(7) as 
$m \rightarrow 0~(1)$ so that as  $m \rightarrow 1,$ four of
the states $\psi_{2,3,4,5}$ must merge at $E = 7$. Besides, it was
also not clear then that in this case there are only two bound bands
followed by a continuum band and that the upper band is a bit unusual
in that both of its band edges are of period $4K$ and inside the band there
are two degenerate states of period $2K$.

(ii) $a - b$ = even integer:

{}From eq. (\ref{2.6a}), it follows that in this case there are
at most ${a+b+2\over 2}$ band gaps of period $2K$.
Explicit examples confirm
that this is indeed so. Hence, in this case one has $a+b+1$
number of nondegenerate states. Quite remarkably, all these
are QES states for which solution can be obtained from
Table III of I when $p = a(a+1), q  =
[a-(a+b)][a-(a+b+1)]$. 

On the other hand, from eq. (\ref{2.7}), it follows that
there are at most ${a - b\over 2}$ band gaps of period $4K$ and
hence $a - b$ number of nondegenerate states of period 4K.
Specific examples confirm the expectation. These are all
QES states which can be obtained from Table III of I in
case $p = a(a+1), q = [a-(a - b)][a-(a - b-1)]$.

Thus when $a - b$ is an even integer (with both $a,b$ integer), then there
are ${a+b+2\over 2}$ band gaps of period $2K$ but only ${a - b\over 2}$
band gaps of period $4K$ and in principle the corresponding band edges are all
analytically known from Table III of I. In view of the fact that the band edge
wave functions in increasing order of energy are of period $2K, 4K, 4K,
2K, 2K,...$, it then follows that in
this case too there must be $b$ zero band gaps of period $4K$
i.e. there must be $b$ doubly degenerate states of period $4K$.
Unfortunately, so far, we have not been able to obtain
analytic solution for even one of these ($2b$) states.
Thus in this case also there are in all $a$  bound bands followed
by a continuum band out of which the top $b$ bound bands are
again bit unusual in that both of their band edges are of 
period $2K$ and two degenerate states of period $4K$ lie
inside each of these bound bands.

As an illustration, consider the case of $p= 12, q = 2$ i.e.
$a = 3, b = 1$. As described in I, this potential is oscillatory 
for $m < \frac{5}{6}$, but has interesting structure coming from 
secondary extrema for $m > \frac{5}{6}$.  From the above discussion 
it follows that the (12,2) potential must have
2 nondegenerate QES states of
period $4K$. Using Table III of I, it is easily seen
that the eigenvalues and the eigenfunctions of these states is
given by
\be\label{2.8}
\psi_1 = \cn \,x ~\dn^2\,x, ~~  E_1 = 1 +4m~,
\ee
\be\label{2.9}
\psi_2 = \sn \,x ~\dn^2\,x~,~~ 
 E_2 = 1+9m~.
\ee
In addition there must be 5 nondegenerate states of period
$2K$ whose eigenvalues and eigenfunctions are easily obtained from Table III 
of I. In particular, the eigenvalues and the eigenfunctions of
two of the states are
\be\label{2.10}
\psi_{8,3} = {\sn \,x~\cn \,x\over \dn \,x} [ 5m
\sn^2\,x-3-m \pm\delta_7]~, ~~ E_{8,3} = 10+2m\pm 2\delta_7~,
\ee
where $\delta_7 = \sqrt{9-9m+m^2}$. On the other hand, the
remaining three eigenstates have the form 
\be\label{2.11}
\psi_{0,4,7} = {1\over \dn \,x} [A+B \sn^2\,x +D \sn^4 \,x ]~,
\ee
and the three corresponding eigenvalues $E_{0,4,7}$ satisfy
the cubic equation (see eqs. (39) and (40) of I)
\be\label{2.12}
\lambda^3 - 4(2m+5)\lambda^2+16(4+11m)\lambda
-192m (2+m) =0~,~~E = \lambda +m~.
\ee
In view of the oscillation theorem, it is then clear that
there must also be two degenerate states
$(\psi_{5,6})$ of period $4K$ whose energy tends to 9(13) as
$m\rightarrow 0(1)$ so that the four states
$\psi_{4,5,6,7}$ must merge at E = 13 as $m\rightarrow 1$.
Thus, as shown in Fig. 2, in this case, there are three bound bands followed by
a continuum band. Further, the upper-most bound band has
both of its band edges of period $2K$ and in between are the
two degenerate states of period $4K$ whose energy eigenvalue and
eigenfunctions are not known analytically. We have 
computed the degenerate energy eigenvalue numerically and it is shown by
a dotted line in Fig. 2. 
\begin{figure}[ht] \label{fig2}
    \centering
    \epsfig{file=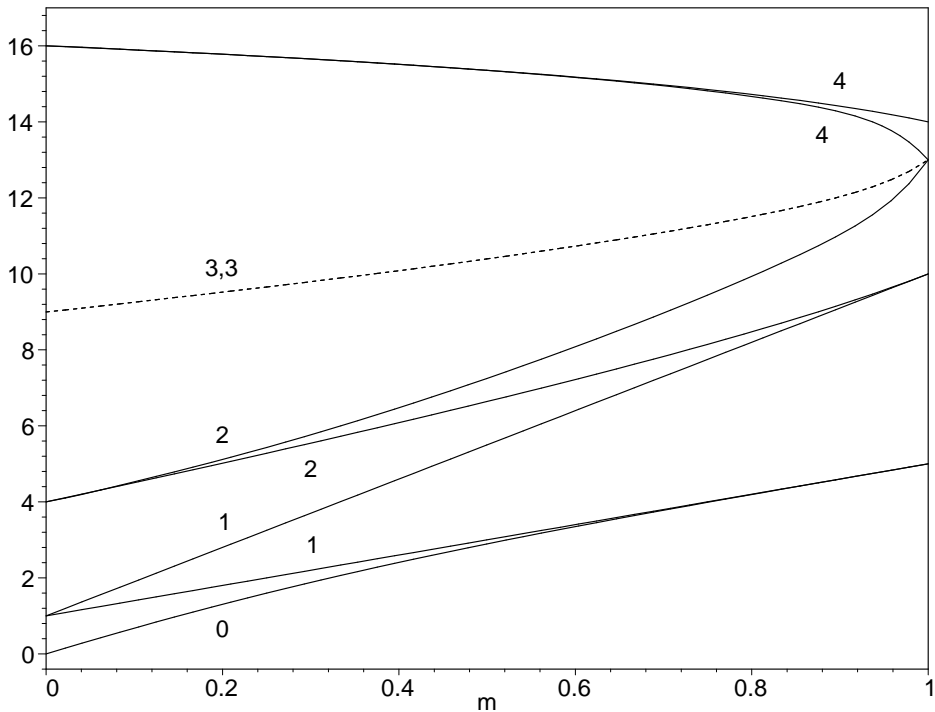,clip=,width=5.0in}
\caption{Band edge energies for the associated Lam\'e potential (12,2)
as a function of the elliptic modulus parameter $m$. The band edges
are labelled by the number of wave function nodes in the interval
$2K(m)$. Note that the band gap between the two states with 3 nodes
is zero. The energy eigenvalue $E_5=E_6$ of these degenerate
states has been calculated numerically and is shown by a dotted line.  }
\end{figure}

What happens if $a = b$ with both being integer? In this
case the associated Lam\'{e} potential has period $K(m)$ rather
than $2K(m)$ and hence the results from Ince's equation are
not directly applicable. However, as has been shown in I
for $a = b$ = 1,2 ( and also explicitly verified for
$a=b=3,4$), in this case there are $a$ bound bands followed by
a continuum band and $2a+1$ band edges all of which are in
principle explicitly known from Table III of I. Further, in
this case, there are no band gaps of zero width.

\sss
{\noindent\bf III. $a,b$ Half Integral and Infinite Number of Band
Gaps of Zero Width}

Let us now specialize to the case when both $a$ and $b$ are
half-integral such that $p > q > 0$. From eqs. (\ref{2.6a})
and (\ref{2.7}) it is clear that in this case either the
roots $\mu_1$ and $\mu_2$ or $\mu_1^{*}$ and $\mu^{*}_2$ are
integral but not the both. In particular, if $a - b$ is an odd integer,
then both the roots $\mu_1$ and $\mu_2$ are integral while
$\mu_1^{*}$ and $\mu^{*}_2$ are not integral while if $a - b$ is
an even integer, then the roots $\mu^*_1$ and $\mu_2^*$ are integral
while $\mu_1$ and $\mu_2$ are not integral. Thus it follows
that unlike the integral $a, b$ case, here one will in general have
infinite number of bands but only finite number of band
gaps of either period $2K$ or $4K$ depending on whether $a - b$ is an odd or an even 
integer. As before, let us discuss
the two cases separately.

(i) $a - b$ = odd integer:

{}From eqs. (\ref{2.6a}) and (\ref{2.7}) it follows that in
this case both $\mu_1$ and $\mu_2$ are integral while
$\mu^*_1$ and $\mu_2^*$ are half integral. Thus in this
case, there will be infinite number of band gaps of period
4K but {\it at most} $~{a+b+2\over 2}$ band gaps of period $2K$.
However, a study of several explicit examples show that
there are in fact only ${a - b+1\over 2}$ number of band gaps
of period $2K$ and hence there are only $a - b$ number of
nondegenerate states of period $2K$ and all these are QES
states. However, using Table III of I it is easily shown that
in this case there are in fact a+b+1 QES states of period
$2K$. This then implies that the remaining $2b+1$ QES
states must correspond to doubly degenerate eigenstates of
period $2K$ i.e. in this case there are $b+{1\over 2}$ doubly
degenerate QES states of period $2K$, each of which is lying inside
a band. It is interesting to note that whereas when both a
and b are integers, then one does not have analytic
expression for even one doubly degenerate state, when both
a,b are half integral, 
energy eigenvalues
and eigenfunctions are known in principle from Table III 
of I for $b+{1\over 2}$ doubly degenerate QES states.

Summarizing, when a - b is an odd integer, then there are infinite
number of bands out of which except for the lowest $a - b$
bands, the rest are rather unusual in that both of their
band edges have period $4K$ and two degenerate states of
period $2K$ reside inside each of these bands.

As an illustration, consider the case of $p = {15\over 4},
q = 3/4$ i.e. $a = 3/2, b = 1/2$. From the above discussion,
it follows that in this case one must have one
nondegenerate and one doubly degenerate QES state of period $2K$.
Using Table III of I it is easily seen that the eigenvalue
and the eigenfunction for the nondegenerate state is given
by 
\be\label{3.1}
\psi_0 = \dn^{3/2} \,x, \ \ E_0 = {9m\over 4}~,
\ee
while the energy eigenvalue and the corresponding two degenerate
eigenfunctions are given by
\be\label{3.2}
\psi_{3} = {\sn \,x~\cn \,x\over
\dn^{1/2} \,x}~, ~ \psi_4 = {[2 \sn^2\,x-1]\over \dn^{1/2}\,x}~,~~E_{3,4} = 4 + {m\over 4}~.
\ee
Note that in this case, out of the infinite number of bands,
except for the lowest band, all other bands have both of their band
edges of period 4K.

Another case, already discussed in I (see Table VII) is
when $p = {63\over 4}, q = 3/4$ i.e. $a = 7/2, \ b = 1/2$.
As shown there, in this case one has three nondegenerate  QES
states of period $2K$ and one doubly degenerate level of the
same period. Further, in this case, but for the lowest three bands,
all other bands have both of their band edges of period 4K.

(ii) $a - b$ = even integer:

{}From eqs. (\ref{2.6a}) and (\ref{2.7}) it follows that in
this case both $\mu^*_1$ and $\mu^*_2$ are integral while
$\mu_1$ and $\mu_2$ are half integral. Hence, in this case
there will be infinite number of band gaps of period $2K$ but
{\it at most} $~{a+b+1\over 2}$ number of band gaps of period
4K. However, a study of several explicit examples reveals
that in this case there are only ${a - b\over 2}$ number of
band gaps and hence a - b number of nondegenerate QES states
of period 4K. However, using Table III of I it is easily
shown that in this case there are in fact a+b+1 number of
QES states of period $4K$. This implies that in addition to
the a - b nondegenerate QES states, there must also be 2b+1
QES degenerate states of period $4K$ i.e. in this case also
there are $b+1/2$ doubly  degenerate QES states of period
$4K$, each of which is lying inside a band.

Thus, in this case too, there are an infinite number of bands
out of which except for the lowest a - b bands, the rest are
rather unusual in that both of their band edges are of
period $2K$ and two degenerate states of period $4K$ reside
inside each of these bands.

As an illustration, consider the case of $p ={35\over 4}, q
= 3/4$ i.e. $a = 5/2, b = 1/2$. In view of the above
discussion, we expect two nondegenerate and one doubly
degenerate QES states of period $4K$. Using Table III of I it is
easily shown that the energy eigenstates for the two 
nondegenerate states are
\be\label{3.3}
\psi_1 = \cn \,x ~\dn^{3/2}\,x~,~E_1 = 1+{9m\over 4}~,
\ee
\be\label{3.4} 
\psi_2 = \sn \,x~\dn^{3/2}\,x~,~E_2 = 1+{25m\over 4}~,
\ee
while for the
two degenerate states one has
\be\label{3.4a}
 \psi_5 = {\cn \,x~(4\sn^2\,x-1)\over
\dn^{1/2}\,x} \ , \ \psi_6 = {\sn \,x\,(4 \sn^2\,x-3)\over
\dn^{1/2}\,x} ~,~E_{5,6} = 9+{m\over 4}~.
\ee
Thus, in this case, (out of the infinite number of bands)
except for the two lowest bands, all other bands are a bit
unusual in that both of their band edge eigenfunctions are
of period $2K$.

We thus have seen that for half integral $a,b$ with $a > b$,
one has $a - b$ nondegenerate and 
$b+{1\over 2}$ doubly degenerate QES states of
period $2K$ or $4K$ depending on whether $a - b$ is an odd or an even
integer. Thus for $b = 1/2, 3/2,\ldots (a > b)$, we expect
1,2,... doubly degenerate QES states. We now show that the
energy eigenvalues of these doubly degenerate states can be
easily obtained analytically in case $b = 1/2$ or $3/2$ and $a$
being arbitrary half integer (with $a > b$). In particular,
if we start with the ansatz
\be\label{3.4b}
y_1(x) = \sum^{N+1}_{k=0} A_{k}~ \sn^{2k}\,x~,
\ee
then on substituting it in the associated Lam\'{e}
eq. (\ref{2.1a}) and equating coefficient of terms
with $\sn^{2N+4} \,x$ and $\sn^{2N+2} \,x$ we find that
in case $a = 2N+3/2, b = 1/2$ then the corresponding
energy eigenvalue is
\be\label{3.7}
E = (2N+2)^2 + m/4, \ N = 0,1,2,...~.
\ee
In fact $E$ as given by eq. (\ref{3.7}) is also the
energy eigenvalue in case we start with the ansatz 
\be\label{3.8}
y_2(x) = \sum^N_{k=0} A_{k} \sn^{2k+1} \,x~ \cn \,x~,
\ee
and substitute it in eq. (\ref{2.1a}). As expected, the
results given in eq. (\ref{3.2}) agree with these given
above in case N = 0 while for N = 1 our results agree with
those given in Table VII of I.

Similarly, it is easily shown that in case $a = 2N + 5/2, b
= 1/2$ then the degenerate eigenvalue and the corresponding
eigenfunctions are
\be\label{3.9}
E = (2N+3)^2 + m/4, \ N = 0,1,2,...~,
\ee
\be\label{3.10}
y_1(x) = \sum^{N+1}_{k=0} \ A_{k} \,\sn^{2k+1} \,x~, \ \ y_2(x)=
\sum^{N+1}_{k=0} B_{k} \,\cn \,x~ \sn^{2k}\,x~.
\ee
For the special case of $N = 0$, our results agree with those
given in eq. (\ref{3.4a}).

On the other hand, for $b = 3/2$ and $a = 2N+ 5/2$, the 
two degenerate energy eigenvalues are
\be\label{3.11}
E = 4N^2+12N+10+{5m\over 4}\pm \sqrt{(4N+6)^2-(4N+6)^2
m+m^2}~, \ N = 0,1,2,...~,
\ee
while the corresponding eigenfunctions are of the form
\be\label{3.12}
y_1(x) = \sum^{N+2}_{k=0} A_{k} \,\sn^{2k} \,x~, \ \ y_2(x) =
\sum^{N+1}_{k=0} B_{k} \, \cn \,x~ \sn^{2k+1}\,x~.
\ee
However, for $b = 3/2$ and $a = 2N+7/2$ , the two
degenerate energy eigenvalues are
\be\label{3.13}
E = 4N^2+16N+17+{5m\over 4}\pm\sqrt{16(N+2)^2-16(N+2)^2
m+m^2}~,~~N=0,1,2,...~,
 \ee
while the corresponding degenerate eigenfunctions are of
the form
\be\label{3.14}
y_1(x) = \sum^{N+2}_{k=0} A_{k} \,\sn \,x^{2k+1}~,~~ y_2(x) =
\sum^{N+2}_{k=0} B_{k} \,\cn \,x ~\sn^{2k} \,x~.
\ee

It is worth adding that 
in case $b = - 1/2 (a>b)$, 
one obtains
$a+{1\over 2}$ nondegenerate QES states of period
$2K(4K)$ depending on if $a+{1\over 2}$ is an odd or an even
integer. Note that when $b=-1/2$ then $q=-1/4$ while $p >0$.
As an illustration, consider $a=5/2, b=-1/2$ i.e. $p=35/4, q=-1/4$.
In this case we have 3 QES states of period $2K$ whose energy 
eigenvalues and eigenfunctions are easily obtained from Table III
of I when $q=(a-2)(a-3)$ and of course $p =a(a+1)$. 

Finally, let us discuss the case when $a,b$ are neither integral nor 
half integral but are such that either $a+b$ or $a - b$ is an integer. 
It is easily seen from eqs. (\ref{2.6a}) and (\ref{2.7}) that if 
either $a+b=2N$ or $a - b=2N+1$ then there are at most $N+1$ band gaps of
period $2K$ while if $a - b=2N$ or $a+b=2N-1$ then there are at most $N$
band gaps of period $4K$. Further, from Table III of I we find that 
when either $a+b=2N$ or $a - b=2N+1$ then there are precisely $2N+1$
QES states of period $2K$ while if $a+b=2N-1$ or $a - b=2N$ then there
are precisely $2N$ QES states of period $4K$ and all these are 
nondegenerate states. 

\sss
{\noindent\bf IV. \ Mid-Band States}

For the Lam\'{e} potential, the majority of results are for integral $a$
(and hence $p$). However, for half-integral values of $a$, analytic expressions for $a+{1\over 2}$ mid-band states (of period $8K$)
have been obtained \cite{ar}. In
particular, it is known that if $ a = 2N+{1\over 2}$, there are $2N+1$
mid-band states of the form ($N=0,1,2,...$)
\be\label{4.1}
\psi(x)= \sqrt{\dn \,x+\cn \,x}~  u(x)~,
\ee
where
\be\label{4.2}
u(x) = \sum^N_{k=0} A_k \sn^{2k} \,x +\sum^{N-1}_{k=0} B_k
~\cn \,x ~\dn \,x ~\sn^{2k}\,x~,
\ee
while if $ a= 2N+{3\over 2}$, then there are $2N+2$ states 
of the form (\ref{4.1}) but where ($N=0,1,2,...$)
\be\label{4.3}
u(x) = \dn \,x~\sum^{N}_{k=0} A_k~ \sn^{2k} \,x +\cn
\,x~\sum^{N}_{k=0} B_k~ \sn^{2k}\,x~.
\ee
Further, since the Lam\'{e} equation is invariant under
$x\rightarrow x+2K(m)$, it follows that in each case one obtains
another solution with the {\it same} energy by changing $\cn \,x$
to  $-\cn\,x$.

In this section we show that the associated Lam\'{e} potential
(2) also has a similar form of mid-band solutions (of
period $8K$) in case $a$ is half integral while $b$ takes
integral values. In particular, if $a = k+{1\over 2}, b =s$
while $k = 2N$ with $N,k,s$ being nonnegative integers, then 
for a given $k$, one obtains $k+1$ doubly degenerate solutions
for every possible (non-negative integral) 
value of $s$. Further, in this case too,
all solutions are also doubly degenerate since the associated
Lam\'{e} equation is invariant under $\cn\,x\rightarrow -
\cn\,x$. 

We begin by substituting the ansatz 
\be\label{4.4}
y(x) = \sqrt{ \dn \,x+\cn \,x} \ z (x)~,
\ee
into eq. (\ref{2.1a}). We find that $z(x)$ satisfies the equation
\bea\label{4.5}
&& \sn \,x~\dn \,xz''(x)+[2bm \sn^2\,x~\cn \,x-\dn \,x
+\cn \,x~\dn^2\,x] z'(x) \nonumber \\
&& +[\lambda_1 \sn \,x ~\dn \,x - rm \sn^3 \,x~ \dn \,x - bm \cn \,x
~\sn \,x+bm \cn^2\,x ~\sn \,x ~\dn \,x] z (x) = 0~,
\eea
where
\be\label{4.6}
\lambda_1 = \lambda - {1+m\over 4}, \ r = (a+1-b) (a+b)-
3/4~. 
\ee
Not surprisingly, $z(x)$ = constant is a solution with energy 
$E = 1+{m\over 4}$ (note $ E = \lambda+mb^2)$ provided $b =
0$ and $a = 1/2$ i.e. $p = 3/4, q=0$. Following the discussion given
above, we try the ansatz (\ref{4.3}) with $N=0$, i.e.
\be\label{4.7}
z(x) = A \,\dn \,x + B \,\cn \,x~,
\ee
in eq. (\ref{4.5}). It is easily shown that there are two
possible solutions in this case

(i) $b = 0, a = 3/2:$

\be\label{4.8}
\psi (x) = [ \dn \,x - (1-m \pm \sqrt {1-m+m^2})\,
\cn \,x]~\sqrt{\dn \,x+\cn \,x}~, ~
E = {5\over 4}(1+m)\pm\sqrt{1-m+m^2}~. 
\ee

(ii) $b = 1, a = 1/2:$

\be\label{4.9}
\psi(x) = [ 1- {2\,\cn \,x\over \dn \,x} ] \sqrt{\dn \,x+ \cn \,x}~,~
 E = {9+m\over 4}~.
\ee
Several comments are in order at this stage.
\begin{enumerate}

\item On making use of the fact that as $m\rightarrow 0, 
\dn\,x \rightarrow 1$ while $\cn \,x\rightarrow \cos \,x$ it is
easily shown that in case $b = 0, a = 3/2$, the two solutions
go over to $\cos {x\over 2}$ and $\cos {3x\over 2}$ with
energies ${1\over 4}$ and 9/4 respectively. On the other
hand, as $m\rightarrow 1$, the two states go over to
the ground and excited states of the potential $V =
{15\over 4} \tanh^2 x$ with eigenvalues ${3\over 2}$ and $7/2$
respectively.

\item On the other hand, as $m\rightarrow 0$ the solution
with $b = 1$ and $a = 1/2$ goes over to $\cos {3x\over 2}$ with
energy ${9\over 4}$ while as $m\rightarrow 1$, 
it goes over to the ground state of the potential
$V={3\over 4} \tanh^2 x+2$ with energy eigenvalue ${5\over 2}$.

\item Degenerate solutions are obtained in all these cases
by changing $\cn \,x$ to -$\cn \,x$, and as $m\rightarrow 0$ these
go over to $\sin {x\over 2}$ or $\sin {3x\over 2}$ as the case may be.

\end{enumerate}

One can now immediately generalize to the general ansatz (\ref{4.3})
and show that for a given
$N$, if $ a = k+1/2$ and $b = s$ with $k = 2N+1-s
~(N=0,1,2,...)$ then $k+1$ doubly degenerate eigenvalues and
eigenfunctions can be obtained for every possible
nonnegative $s$. For example, let us consider the ansatz
(\ref{4.3}) with $N=1$, i.e. 
\be\label{4.10}
z(x) = \dn \,x~ [A_0+A_1 \sn^2 \,x]+\cn \,x~ [B_0+B_1 \sn^2\,x]~.
\ee
After substituting this ansatz in eq. (\ref{4.5}) and performing
lengthy algebraic manipulations, it is easily shown that there are four
possible solutions in this case.

(i) $b = 0, a = 7/2:$

As is well known \cite{ar}, in this case  
$\lambda_1 = E - \frac{1+m}{4}$
satisfies a quartic equation
\be
\lambda_1^4 -20(1+m)\lambda_1^3 +18(6+19m+6m^2)\lambda_1^2
-36(4+39m+39m^2+4m^3)\lambda_1 +135m(8+23m+8m^2) = 0~,
\ee
all of whose roots are real
and that as $m\rightarrow 0$ the solutions go over
to $\cos {x\over 2}, \cos {3x\over 2}, \cos {5x\over 2}, \cos
{7x\over 2}$ with energies ${1\over 4}, {9\over 4},
{25\over 4}$ and ${49\over 4}$ respectively.

(ii) $b = 1, a = 5/2:$

In this case $\lambda_1$ can be shown to satisfy the cubic
equation (note $E = \lambda_1+{1+m\over 4}+mb^2)$
\be\label{4.11}
\lambda^3_1 - (5m+14)\lambda^2_1 +
(24+88m-m^2)\lambda_1+5m^3-98m^2-96m = 0~,
\ee
whose all three roots are real for any $m~(0\leq m\leq 1)$.
As $m\rightarrow 0$, we find that the solutions go over to
$\cos {x\over 2}, \cos {3x\over 3}, \cos {7x\over 2}$ with
energies 1/4, 9/4, 49/4 respectively.

(iii) $b = 2, a = 3/2:$

In this case there are two solutions with the corresponding energies being
\be\label{4.12}
E = {29+5m\over 4}\pm \sqrt{25-25m+m^2}~.
\ee
Note that as $m\rightarrow 0$ the two energies go over to 
${9\over 4}$ and ${49\over 4}$ and the corresponding
solutions go over to $\cos {3x\over 2}$ and $\cos {7x\over 2}$
respectively. 

(iv) $b = 3, a = 1/2:$

In this case there is only one solution given by 
\be\label{4.13}
\psi = \bigg (\dn \,x [ 1-{(4-m)\over 3}
\sn^2\,x] - {4\over 3} \cn \,x~ [ 1-(2-m) \sn^2\,x] \bigg )
\sqrt{\dn \,x + \cn \,x}~,~ E = {49+m\over 4}~.
\ee
It is easily checked that as $m\rightarrow 0$ the
solution goes over to $\cos {7x\over 2}$ and the
corresponding energy is ${49\over 4}$.

On the other hand if $a = k+{1\over 2}, b = s$ and $k = 2N-s
~(N=0,1,2,...$) then we start with the ansatz (\ref{4.2})
and obtain $k+1$ (doubly degenerate) eigenvalues and
eigenfunctions for every possible nonnegative $s$. For $N = 0$,
the only possibility is of course $a = 1/2, b = 0$ and in
this case the solution is already well known \cite{ar}. For $N = 1$ we
start with the ansatz,
\be\label{4.14}
z(x) = A_0 + A_1~ \sn^2 \,x + B_0~ \cn \,x ~\dn \,x~.
\ee
On substituting this ansatz in eq. (\ref{4.5}), after
lengthy but straightforward algebraic manipulations, we find the following three solutions: 

(i) $b = 0, a = 5/2:$

As is well known \cite{ar}, in this case  
$\lambda_1 = E - \frac{1+m}{4}$
satisfies a cubic equation 
\be
\lambda_1^3 -8(1+m)\lambda_1^2 +4(3+13m+3m^2)\lambda_1 -48m(1+m) =0~,
\ee
all of whose roots are real
and as $m\rightarrow 0$, the
solutions go over to $\cos {x\over 2}, \cos {3x\over 2},
\cos {5x\over 2}$ with energies ${1\over 4}, {9\over
4}$ and ${25\over 4}$ respectively.

(ii) $b = 1, a = 3/2:$

In this case there are two solutions with the corresponding energies being
\be\label{4.15}
E = {13+5m\over 4}\pm \sqrt{9-9m+m^2}~.
\ee
Note that as $m\rightarrow 0$ the energies go over to
${1\over 4}$ and ${25\over 4}$ while the corresponding solutions go
over to $cos {x\over 2}$ and $\cos {5x\over 2}$ respectively.

(iii) $b = 2, a = 1/2:$

In this case there is only one solution
\be\label{4.16}
\psi = \sqrt{\dn \,x\,+\,\cn \,x} \bigg[
1-{(4-m)\over 3} \sn^2 \,x - {2\over 3} \cn \,x ~\dn \,x\bigg ]~,~E = {25+m\over 4}~, 
\ee
As $m\rightarrow 0$ the solution goes over to
$\cos{5x\over 2}$ with energy 25/4.

Before closing this section it is worth pointing out that the mid-band states
have already 
been obtained in I in case $a=b= N+1/2, (N= 0,1,2,...)$. In particular,
it may be noted that when $a=b$, then the associated Lam\'{e} potential has
period $K$ rather than $2K$ and hence the band edges will be of period $K$
and $2K$ while the mid-band states will be of period $4K$. Now if one looks
at the Table III of I then one notices that if $a=b=N+1/2$, then there are
$N+1$ doubly degenerate QES states of period $4K$ which 
are obtained in principle 
from the Table III of I in case $a=N+1/2$ and  
$q =[a-(N+1)][a-(N+2)]$ . For example, for $N=0$,
i.e. for $p=q=3/4$, the 
doubly degenerate mid-band states are
\be\label{4.17}
\psi_1 = {\cn \,x\over \sqrt{\dn \,x}}~,
~\psi_2 = {\sn \,x\over \sqrt{\dn \,x}}~,~E = 1+{m\over 4}~.
\ee
On the other hand, for $N=1$ the pair of doubly degenerate 
mid-band states are
\be
\psi_1 = \frac{\cn \,x (4-m-2\sn^2 \,x)}{\dn^{3/2} \,x}~,
~\psi_2 = \frac{\sn \,x (4-2\sn^2 \,x)}{\dn^{3/2} \,x}~,~E= 1+{9m\over 4}~,
\ee 
\be
\psi_1 = \frac{\cn \,x ~\sn^2 \,x}{\dn^{3/2} \,x}~,
~\psi_2 = \frac{\sn \,x (2\sn^2 \,x-1)}{\dn^{3/2} \,x}~,~E= 9+{m\over 4}~.
\ee

\sss
{\noindent\bf V. Conclusion and Open Problems:}

In this paper we have clarified several issues regarding the associated 
Lam\'{e} potential. In particular, we have shown that when both $a,b$ are
integers then just like Lam\'{e}, the associated Lam\'{e} potential is
also a finite band problem. The only difference from the Lam\'{e} case
arises when $a \ne b$ - in that case one has some bands with both
band edges of the same period. We have also seen that when both $a,b$
take half integral but unequal values then one has a genuine QES problem and
band edges of either period $2K$ or $4K$ are known, but not both. We have also
shown that in this case, but for the few low lying bands, all other bands
have both of their band edges of the same period ($2K$ or $4K$).
Finally, when $a$ is  a half integer and $b$ is an integer, we can 
obtain several mid-band states.

It would be nice if one could 
(i) say something concrete about the band structure 
when $a,b$ are
neither integral or half integral; (ii) derive
dispersion relations for at least some of the finite band associated
Lam\'{e} problems; (iii) obtain
the band edges of the associated Lam\'{e} problem algebraically, analogous to the Lam\'{e} potential. We hope to
address some of these issues in the near future. \\

\end{document}